\documentclass[draftclsnofoot,onecolumn,12pt]{IEEEtran}
\usepackage{graphicx}
\usepackage{epstopdf}
\usepackage{subfigure}
\usepackage{booktabs}
\usepackage{caption}
\usepackage{color}
\usepackage{bm}
\usepackage{CJK}
\usepackage{indentfirst}
\usepackage{amsmath}
\usepackage{amssymb}
\usepackage{float}
\usepackage{multirow}
\usepackage{algorithm}
\usepackage{algpseudocode}
\usepackage{amsmath}
\usepackage{flushend}

\usepackage{geometry}
\newtheorem{theorem}{Theorem}

\begin{document}
\title{Angle-Dependent Phase Shifter Model for Reconfigurable Intelligent Surfaces: Does the Angle-Reciprocity Hold?}
\author{\IEEEauthorblockN{Weicong~Chen,~Lin~Bai,~Wankai~Tang,~Shi~Jin,~Wei~Xiang~Jiang, and~Tie~Jun~Cui}\\
\thanks{{W. Chen, W. Tang, and S. Jin are with the National Mobile Communications Research Laboratory, Southeast University,
Nanjing, China (e-mail: cwc@seu.edu.cn; tangwk@seu.edu.cn; jinshi@seu.edu.cn)}\par
{L. Bai, W. X. Jiang, and T. J. Cui are with the State Key Laboratory of Millimeter Waves, Southeast University,
Nanjing, China (e-mail: nustbl@163.com; wxjiang81@seu.edu.cn; tjcui@seu.edu.cn)}}
}

\maketitle
\begin{abstract}
The existing phase shifter models adopted for reconfigurable intelligent surfaces (RISs) have ignored the electromagnetic (EM) waves propagation behavior, thus cannot reveal practical effects of RIS on wireless communication systems. Based on the equivalent circuit, this paper introduces an angle-dependent phase shifter model for varactor-based RISs. {To the best of our knowledge, this is the first phase shifter model which reveals that the incident angle of EM waves has influence on the reflection coefficient of RIS}. In addition, the angle-reciprocity on RIS is investigated and further proved to be tenable when the reflection phase difference of adjacent RIS unit cells is invariant for an impinging EM wave and its reverse incident one. The {angle-dependent characteristic} of RIS is verified through full-wave simulation. According to our analysis and the simulation results, we find that the angle-reciprocity of varactor-based RIS only holds under small incident angles of both forward and reverse incident EM waves, thus limits the channel reciprocity in RIS-assisted TDD systems.

\end{abstract}

\section{Introduction}
{Extended from metasurface, the introduction of RISs, or intelligent reflecting surfaces (IRSs), has the potential to innovate communication architecture \cite{T. J. Cui}\cite{J. Zhao} and customize smart radio environment \cite{M. Di Renzo} for next generation wireless communication systems.} Due to their brilliant abilities in manipulating EM waves with low-cost hardware, RISs have recently received significant attentions. There are already a few recent researches focus on RIS from a variety of points, such as ergodic spectral efficiency analysis \cite{Y. Han}, energy efficiency maximization \cite{C. Huang}, channel estimation \cite{CS_DL}, passive beamforming \cite{Q. Wu} and implementation \cite{W.Tang_EL}\cite{W. Tang_TWC}. It has been shown that the RIS¡¯s phase
control combined with conventional transmission control can bring additional performance gain compared with traditional
wireless systems without RIS \cite{S. Gong}. When RISs are designed as reflecting array that work in the middle of the channel between the transmitter and the receiver, their models are worth investigating to see how RISs change the channel. \par

Phase shifter models are broadly adopted for elements of RIS. They can be classified into three categories according to the reconfigurability of RIS's reflection amplitude and phase. The first one is adopted in most of existing works where phase shifters were assumed to be constant modulus with continuously adjustable phase \cite{Y. Han}-\cite{CS_DL}. In the second one, the amplitude and phase of the phase shifter were considered to be independently adjustable \cite{Q. Wu}\cite{amp_each}. The third one is the amplitude-phase dependent phase shifter model firstly proposed in \cite{amp_pha_dep}.\par
The aforementioned works only concerned about the amplitude and phase response of the phase shifter model but ignored the influence brought by the propagation behavior of EM waves. In such models, RISs generate the same response for EM waves from different directions and they will not break the channel reciprocity in TDD systems. While through experimental results, \cite{pass_loss} reported that the phase response of RIS is sensitive to the incident angle of EM waves due to the spatial dispersion in existing RIS unit cell. Since EM waves are ubiquitous in all directions in actual radio environment, a more accurate and practical phase shifter model is desperately necessary and the impact of RIS on wireless channel should be reviewed.\par

In this paper, we consider RIS-assisted wireless communication systems and provide the signal model on RIS. We then introduce an angle-dependent phase shifter model for the unit cell of varactor-based RIS from its equivalent circuit. After that, we investigate the angle-reciprocity of EM waves on RIS. It is proved that when the reflection phase difference of adjacent unit cells is invariant for an impinging EM wave and its reverse incident one, the propagation direction on RIS can be reversed. Full-wave simulation results verify the {angle-dependent characteristic} of our phase shifter model and demonstrate that the angle-reciprocity can only hold when the forward and reverse incident angles are small.
%

\section{System model}

\begin{figure}
  \centering
  \includegraphics[width=0.7\textwidth]{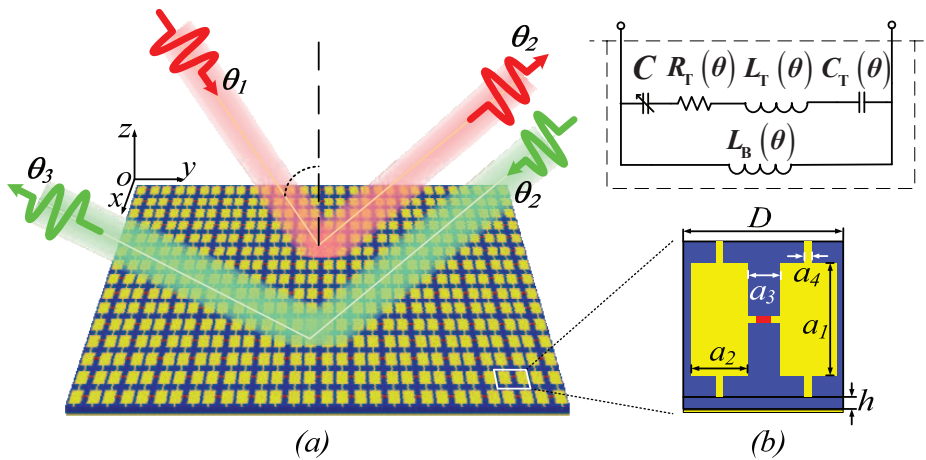}\\
  \caption{($a$) Conceptual illustration of the RIS. ($b$) The schematic of unit cell and its simplified equivalent circuit model.}\label{Fig.RIS}
\end{figure}
We consider RIS-assisted communication systems and introduce the incident and reflected signal model on RIS. Fig. \ref{Fig.RIS}($a$) illustrates a typical RIS that manipulates the propagation direction of EM waves\footnote{{In this paper, EM waves are considered to be plane wave for analyzing the angle-reciprocity. Considering the sub-wavelength of unit cell, this is reasonable since the transmitter/receiver is nearly always at the far-field of the unit cell.}}. The RIS, composed of $N$ columns and $M$ rows unit cells, is placed on the $xoy$ plane. We assume that the RIS unit cells in each column are connected to the same bias network and each columns of the RIS can be individually controlled by the bias voltage. Therefore, incident EM waves in the $yoz$ plane can be reflected to another direction via changing the bias voltages of each column to manipulate the phase and amplitude response of the unit cells. The reflected signal of the incident EM waves is given as
\begin{equation}\label{eq:y}
  {\bf{y}} = {{\bf{\Gamma }}}{\bf{a}_{\rm in}}\left( {{\theta}} \right)
\end{equation}
where ${\bf{a}_{\rm in}}\left( {{\theta }} \right) = {\left[
1,{{e^{j\frac{{2\pi D}}{\lambda }\sin \left( {{\theta }} \right)}}}, \cdots ,{{e^{j\left( {N - 1} \right)\frac{{2\pi D}}{\lambda }\sin \left( {{\theta}} \right)}}} \right]^T}$ is the array response vector of an EM wave, $\theta$ is the incident angle between EM wave propagation direction and the normal of RIS, $D$ is the periodic length of the unit cell, $\lambda$ is the wavelength and ${{\bf{\Gamma }}} = {\rm{diag}}\{ \Gamma_1, \cdots ,\Gamma_N\}  $ is the reflection coefficient matrix of RIS where $\Gamma_n$ is the reflection coefficient of the $n$-th RIS column.\par
The reflection coefficient $\Gamma$ is generally modeled as phase shifter that is independent of EM waves incident angle. However, the experimental results in \cite{pass_loss} revel that the phase of reflection coefficient is related to the incident angle. Therefore, we will introduce an angle-dependent phase shifter model to describe the effect of incident angle on the reflection coefficient of RIS. Furthermore, the angle-reciprocity\footnote{The angle-reciprocity means that when an EM wave impinges RIS from $\theta_1$ and reflected to $\theta_2$, the propagation direction can be reversed without changing the bias networks.} on RIS is investigated, since it is essential for realizing channel reciprocity in TDD systems.

\section{Anlge dependence of RIS}
In this section, the angle-dependent phase shifter is modeled from the equivalent circuit of RIS unit cell, and further utilized to analyze the angle-reciprocity of EM waves on RIS.
\subsection{Angle-dependent phase shifter model}
The reflection-type RIS unit cell is widely modeled as phase shifter, and its specific characteristic is determined by control method and physical structure. According to the control method, the RIS unit cell can be divided into two categories. One is the PIN diode based unit cell that has limited tunable states. The other is varactor diode based and continuously adjustable.\par

In this work, we design a varactor-based reflection-type\footnote{{The PIN-based unit cell has limited states, which can be seen as the discrete case of varactor-based unit cell.}} unit cell with concrete physical structure to investigate the phase shifter model for RIS. The unit cell consists three layers as shown in the bottom of Fig. \ref{Fig.RIS} ($b$). The top layer has two symmetrical metallic patches bridged by a varactor diode while the bottom layer is a metallic ground. The middle layer separates the top and bottom layer with a dielectric substrate. According to \cite{360}\cite{W. Tang_1}, the simplified equivalent circuit of unit cell can be modeled as shown in the top of Fig. \ref{Fig.RIS} ($b$). The impedance of such a parallel resonant circuit is given by

\begin{equation}\label{eq:Z_0}
  Z( {\theta ,f,C} )=\frac{{jf{L_{\rm B}}( \theta  )( {{R_{\rm T}}( \theta  ) + jf{L_{\rm T}}( \theta  ) + \frac{1}{{ {jf{C_{\rm T}}( \theta  )} }} + \frac{1}{{ {jfC} }}} )}}{{jf{L_{\rm B}}( \theta  ) + ( {{R_{\rm T}}( \theta  ) + jf{L_{\rm T}}( \theta  ) +  \frac{1}{{ {jf{C_{\rm T}}( \theta  )} }} + \frac{1}{{ {jfC} }}} )}},
\end{equation}
where $C$ is the capacitance of varactor, $\theta$ and $f$ are the incident angle and frequency of EM waves, respectively. Different from \cite{360}\cite{W. Tang_1}, in \eqref{eq:Z_0}, the equivalent inductance $L_{\rm B}\left( \theta  \right)$ of the bottom layer, the equivalent resistance $R_{\rm T}\left( \theta \right)$, inductance $L_{\rm T}\left( \theta  \right)$ and capacitance $C_{\rm T}\left( \theta  \right)$ of the top layer are modeled to be the function of incident angle, and their specific expressions are determined by geometric parameters of RIS unit cell. This is due to the fact that the equivalent circuit parameters are determined by projected geometric parameters of unit cell. When EM waves illuminate RIS from different directions, some projected geometric parameters of unit cell will change, resulting in variable equivalent circuit parameters.       \par

The reflection coefficient of a reflection-type unit cell is a parameter that describes the fraction of EM waves reflected by an impedance discontinuity in transmission medium \cite{W. Tang_1}. Given \eqref{eq:Z_0} and the impedance of free space $Z_0$, the reflection coefficient, or the phase shifter, of the unit cell is expressed as \cite{D. M. Pozar}
\begin{equation}\label{eq:Gamma}
  \Gamma \left( {\theta ,f,C} \right) = \frac{{Z\left( {\theta ,f,C} \right) - {Z_0}}}{{Z\left( {\theta ,f,C} \right) + {Z_0}}}
\end{equation}
According to \eqref{eq:Gamma}, the phase shifter is modeled as angle-dependent since it is also the function of incident angle $\theta$ different from existing works. {Specially, when \eqref{eq:Gamma} keeps constant for a given incident angle region, it reduces to amplitude-phase dependent phase shifter model in \cite{amp_pha_dep}.}\par

\subsection{Angle-reciprocity analysis}\label{sec:3}
Introducing RISs into wireless communication systems can customize a smart radio environment, meanwhile changes the wireless channel characteristic due to its angle-dependent response. For example, channel reciprocity is the golden rule in traditional TDD systems. However, when the angle-reciprocity on RIS fails, the channel reciprocity in RIS-assisted TDD systems will be broken. \par

To have a direct cognition on the angle-reciprocity on RIS, we provide the following Theorem.

\begin{theorem}\label{Theo-1}
In RIS-assisted wireless communication systems, without modifying the RIS reflection coefficients that reflect the impinging EM waves from $\theta_1$ toward $\theta_2$, the reverse EM waves that incident RIS from $\theta_2$ will be reflected toward $\theta_3$ expressed by
\begin{equation}\label{eq:theta_3_1}
{\theta _3} = \arcsin \left\{ {\sin \left( {{\theta _1}} \right) + \frac{\lambda }{{2\pi D}}\left( {\Delta {\Phi _{{\theta _1},f}} - \Delta {\Phi _{{\theta _2},f}}} \right)} \right\},
\end{equation}
where $\Delta {\Phi _{{\theta _1},f}}$ and $\Delta {\Phi _{{\theta _2},f}}$ are the reflection phase difference of adjacent RIS unit cells for EM waves incident from $\theta_1$ and $\theta_2$, respectively.
\begin{IEEEproof}
See Appendix \ref{App:A}.
\end{IEEEproof}

\end{theorem}\par


\textit{Theorem} \ref{Theo-1} reveals that the angle-reciprocity of EM waves on RIS holds when $\Delta {\Phi _{{\theta_1},f}}- \Delta {\Phi _{{\theta_2},f}}=0$. Under the existing phase shifter models that ignore the propagation behavior of EM waves, $\Delta {\Phi _{{\theta},f}}$ is a constant for arbitrary $\theta$, thus $\theta_3=\theta_1$ is always tenable. However, the introduced phase shifter model is supposed to be angle-dependent, therefore $\Delta {\Phi _{{\theta},f}}$ varies with $\theta$. We know the Taylor series expansion of $f\left(x\right)=\arcsin\left(x\right)$ at $x_0$ can be expressed as
\begin{equation}\label{eq:Theo-2}
  f\left(x\right) = \arcsin\left(x_0\right)+\frac{1}{\sqrt{1-x_0^2}}\left(x-x_0\right) +\mathcal{O}\left(x-x_0\right)
\end{equation}
Denoting $x=x_0 +\delta$ where $x_0=\sin \left( {{\theta _1}} \right)$ and
\begin{equation}\label{eq:delta}
  \delta= \frac{\lambda }{{2\pi D}}\left( {\Delta {\Phi _{{\theta _1},f}} - \Delta {\Phi _{{\theta _2},f}}} \right),
\end{equation}
\eqref{eq:theta_3_1} can be rewriten as
\begin{equation}\label{eq:Theo-1}
  \theta_3 = \theta_1 + \frac{1}{\cos\left(\theta_1\right)}\delta+\mathcal{O}\left(\delta\right)
\end{equation}
It can be seen from \eqref{eq:Theo-1} that when $\delta \ne 0$, the reflected angle of reverse EM waves, $\theta_3$, is also determined by $\frac{1}{\cos(\theta_1)}$. In other words, increasing the forward incident angle $\theta_1$ will expand the deviation between $\theta_3$ and $\theta_1$.\par

At this point, we know that the {angle-dependent characteristic} of RIS determines its angle-reciprocity. More importantly, it will pose new challenges for applications of RIS in practical communication systems. For example, passive beamforming optimization in multipath channel will be more knotty when the angle-dependent phase shifter model is adopted, since signals from different path will have different response on the same RIS unit cell. Furthermore, once the angle-reciprocity on RIS fails, the broken channel reciprocity in TDD system invalidates the conventional channel estimation strategies, and the transmission directions assisted by RIS in uplink and downlink can not be aligned. Therefore, the multifarious system performances obtained by RIS should be reappraised. To ease the effects brought by the {angle-dependent characteristic} of RIS, the partial angle insusceptibility design for RIS is worth investigating, although it is impossible to design a two dimensional structure unit cell with omnidirectional
invariable EM response.

In the following, we will perform full-wave simulation to evaluate the {angle-dependent characteristic} of RIS, since the introduced phase shifter model is determined by the concrete geometric parameters design of the varactor-base RIS unit cell and has no closed-form expression.

\section{NUMERICAL results}
In this section, the full-wave simulation is carried out by using commercial software CST Microwave Studio 2017 (www.cst.com/products/cstmws) to demonstrate the {angle-dependent characteristic} of the varactor-based unit cell and the angle-reciprocity of EM wave on RIS.\par
The structure of unit cell determines its function and characteristic. As shown on the bottom of Fig. \ref{Fig.RIS} ($b$), a reflection-type square unit cell is designed with periodic length $D=8$ mm. The geometric parameters on the top layer are set to $a_1 = 5.5$ mm, $a_2=2.9$ mm, $a_3=1.5$ mm and $a_4=0.4$ mm. The middle layer is a $h=2$ mm thick F4B. A whole metallic ground constitutes the bottom layer. To realize reconfiguration of the RIS, a Skyworks SMV1405-079LF varactor is selected to bridge the patches on the top layer. With the capacitance of varactor varies from maximal $2.67$ pF to minimal $0.63$ pF when the corresponding reverse voltage increasing from $0$ V to $30$ V, the reflection coefficient of unit cell can be continuously tuned.\par

\subsection{Verification of the angle-dependent phase shifter}

To demonstrate the {angle-dependent characteristic}, we set the capacitance of varactor as $0.63$ pF when EM waves impinge the RIS from $0^\circ$, $30^\circ$ and $40^\circ$, respectively. \par

\begin{figure}
  \centering
  \includegraphics[width=0.7\textwidth]{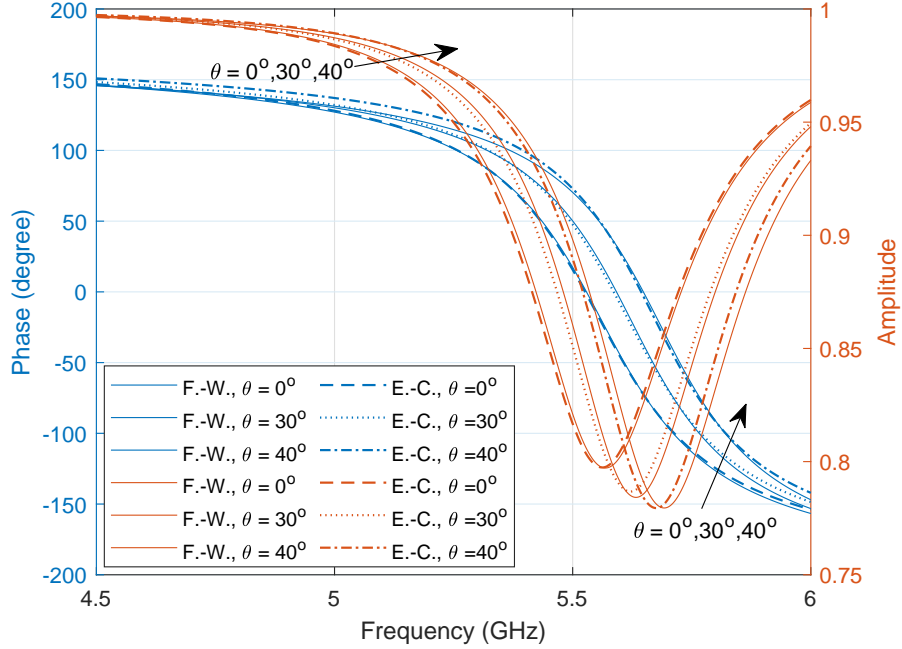}\\
  \caption{Phase and amplitude versus carrier frequency under full-wave (F.-W.) simulation and equivalent-circuit (E.-C.), for different incident angle $\theta$.}\label{Fig.dif_angle}
\end{figure}

Solid lines in Fig. \ref{Fig.dif_angle} demonstrate the characteristics of the phase and amplitude of reflection coefficient under full-wave simulation, for different incident angles. It is observed that the phase and amplitude both vary with the frequency of incident EM wave. We know that when the frequency of incident EM wave approaches to resonance frequency of the unit cell, the phase and amplitude of reflection coefficient will approach to zero and minimum, respectively. {In Fig. \ref{Fig.dif_angle}, it can be seen that as the incident angle of EM waves increases, the phase and amplitude of reflection coefficient changes. Moreover, the zero phase and minimum amplitude will shift to higher frequency.} This implies that resonance frequency of the unit cell increases with the incident angle. According to the simplified equivalent circuit shown in the top of Fig. \ref{Fig.RIS} ($b$), the resonance frequency of the unit cell is $  f_{\rm r}(\theta) = {1}/{\sqrt{\left(L_{\rm B}(\theta)+L_{\rm T}(\theta)\right)\frac{C_{\rm T}(\theta) C}{C_{\rm T}(\theta)+C}}}$. Now that the resonance frequency varies with the incident angle, the equivalent circuit parameters of the phase shifter must be angle-dependent. We set the equivalent circuit parameters of the phase shifter for different incident angles as TABLE \ref{tab:parameters}, and then plot the phase and amplitude of the phase shifter model, i.e., $\Phi \left( {\theta ,f,C} \right)={\rm{arg}}\angle {\Gamma \left( {\theta ,f,C} \right) }$ and $A \left( {\theta ,f,C} \right)=|\Gamma \left( {\theta ,f,C} \right)| $, as the dash lines in Fig. \ref{Fig.dif_angle}. Small gaps between the solid and dash lines in Fig. \ref{Fig.dif_angle} reveal that the phase shifter modeled from the equivalent circuit matches well with the full-wave simulated reflection coefficient of the unit cell. Furthermore, the values of equivalent circuit parameters under different incident EM wave angles prove that the phase shifter is angle-dependent in practical.
\begin{table}[!htb]
\caption{Equivalent circuit parameters of the phase shifter for different incident angles}
\label{tab:parameters}
\centering
\begin{tabular}{|c|c|c|c|}
    \hline
      & $\theta = 0^\circ$ & $\theta = 30^\circ$ & $\theta = 40^\circ$\\
    \hline
    $L_{\rm B}(\theta)$ (nH) & $15.83$ & $15.56$ & $14.44$ \\
    \hline
    $L_{\rm T}(\theta)$ (nH) & $38.26$ & $38.92$ & $35.56$ \\
    \hline
    $R_{\rm T}(\theta)$ ($\Omega$) & $2.20$ & $2.23$ & $2.11$ \\
    \hline
    $C_{\rm T}(\theta)$ (pF) & $15.6$ & $8.9$ & $200$ \\
    \hline
    $f_{\rm r}(\theta)$ (GHz) & $5.53$ & $5.59$ & $5.64$ \\
    \hline

\end{tabular}
\end{table}
\subsection{Evaluation of the {angle-dependent characteristic} of RIS}
Next, we will demonstrate how angle-dependent the varactor-based RIS is, and then evaluate the angle-reciprocity of EM waves on RIS at the operating frequency.\par

\begin{figure}
  \centering
  \includegraphics[width=0.7\textwidth]{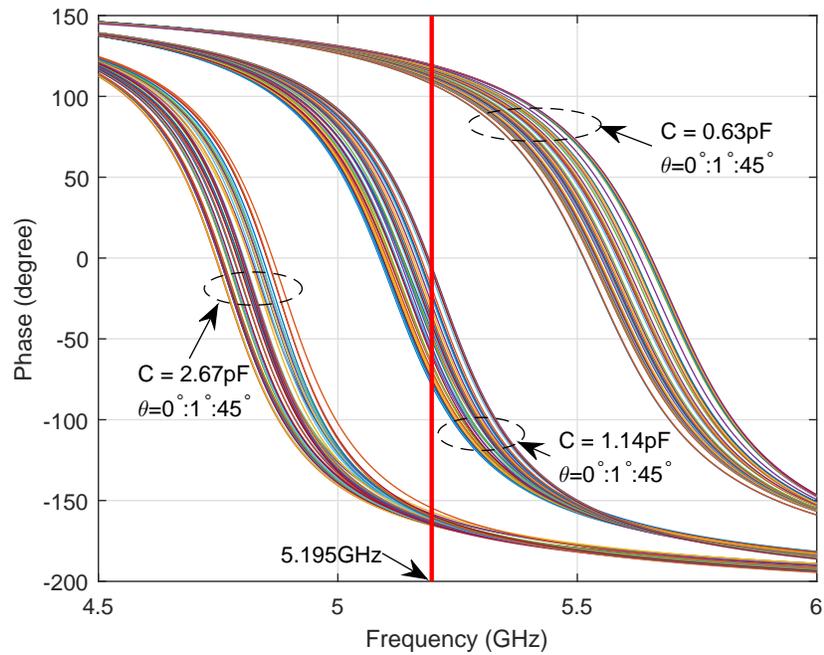}\\
  \caption{Phase of the reflection coefficient versus carrier frequency for incident angles increased from $0^\circ$ to $45^\circ$ step by $1^\circ$.}\label{Fig.three_capa}
\end{figure}

To efficiently manipulate EM waves, when the capacitance of varactor increases from minimum to maximum, the reconfigurable phase range should be as large as possible at the operating frequency of RIS. Fig. \ref{Fig.three_capa} demonstrates the phase of reflection coefficient versus carrier frequency, for incident angles increased from $0^\circ$ to $45^\circ$ step by $1^\circ$, and capacitance of varactor is set to $0.63$ pF, $1.14$ pF and $2.67$ pF, respectively. As shown Fig. \ref{Fig.three_capa}, $f_c = 5.195$ GHz is the frequency that offers largest reconfigurable phase range when incident angle is fixed, thus can be set as the operating frequency of the RIS designed in this paper. Furthermore, it is observed that each bunch of curves has a narrower phase variation range for different incident angles when the slope of curve is smaller. While the phase response becomes more susceptible to the incident angle when the slope of curve is increasing. Moreover, the bunch of curves moves to lower frequency on the whole as the capacitance of varactor tuned larger. These imply that at the operating frequency, the reflection phase in the middle of reconfigurable range is inevitably more susceptible to the incident angle.



\begin{figure}
  \centering
  \includegraphics[width=0.7\textwidth]{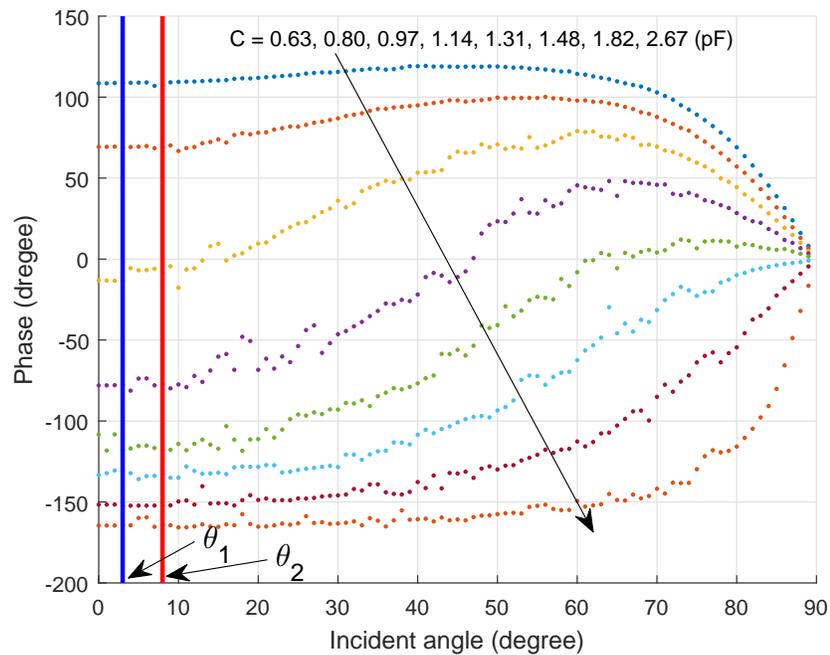}\\
  \caption{Phase of the reflection coefficient versus incident angle at operating frequency $f_c$.}\label{Fig.eight_capa}
\end{figure}
To further demonstrate the angel-dependence, Fig. \ref{Fig.eight_capa} shows the reflection phase versus incident angle at operating frequency $f_c$, for different capacitance. The reconfigurability of the varactor-based RIS is determined by the tunable phase range, which is the gap between the curves determined by $C=0.63$pF and $2.67$pF, respectively, in Fig. \ref{Fig.eight_capa}. As illustrated in Fig. \ref{Fig.eight_capa}, although the tunable phase range remains wide for the incident angle less than $60^\circ$, it will shrink sharply to zero when the incident angle continues to grow. This trend means that the reconfigurability of the varactor-based RIS is deteriorated for an EM wave illuminates from large angle. \par
{In addition}, the gaps (phase differences) between any two curves in Fig. \ref{Fig.eight_capa} roughly remain constant only for incident angle between $0^\circ$ and $10^\circ$. From \textit{Theorem} \ref{Theo-1} we know that constant phase difference $\Delta\Phi_{\theta,f}$ makes $\theta_3=\theta_1$. {In other words, the angle-reciprocity of varactor-based RIS will holds under limited conditions that the incident and reflected angle, $\theta_1$ and $\theta_2$, are both small, for example, $\theta_1$, $\theta_2 \in [-10^\circ,10^\circ]$ in this paper. This angle-reciprocity range may be different for different unit cell design, and the negative range is due to the symmetry of the unit cell.} While increasing the incident or reflected angel of EM wave, the propagation direction cannot be reversed without changing the bias networks of the RIS since $\delta=0$ cannot be guaranteed.

\section{Conclusion}\label{sec:5}
In this paper, we have provided an angle-dependent phase shifter model for the varactor-based RIS in RIS-assisted wireless communication systems. The angle-dependent phase shifter model was introduced based on the equivalent circuit of varactor-based RIS unit cell. Different from the existing phase shifter models, our model considered the effects of oblique EM wave incident angle on reflection coefficient of the varactor-based RIS. Based on the introduced model, the angle-reciprocity on the varactor-based RIS was investigated. It was proved that when the reflection phase difference of adjacent unit cells is invariant for an impinging EM wave and its reverse incident one, the propagation direction on RIS can be reversed. The full-wave simulation results verified the {angle-dependent characteristic} of the varactor-based RIS and further revealed that when the incident and reflected angles of EM waves were small, RIS will be reciprocal according to \textit{Theorem} \ref{Theo-1}. {Otherwise the angle-reciprocity of varactor-based RIS failed, resulting in the break of channel reciprocity.} Therefore, the angle insusceptibility design for RIS is the most vital solution to overcome the challenges in RIS-assisted TDD wireless communication systems. Before that, it is more reasonable to deploy RIS in FDD systems.

%
\appendices
\section{}\label{App:A}
For angle-reciprocity analysis, we only consider the reflection phase $\Phi \left( {\theta ,f,C} \right)={\rm{arg}}\angle {\Gamma \left( {\theta ,f,C} \right) }$, and assume reflection amplitude $|\Gamma \left( {\theta ,f,C} \right)|=1 $ since it is easy to prove that the reflection amplitudes will not impact the direction of reflected EM waves. Considering an EM wave impinges RIS from $\theta_1$ and reflected to $\theta_2$ as shown in Fig. \ref{Fig.RIS} ($a$), the signal model in \eqref{eq:y} can be rewritten as
\begin{equation}\label{eq:in_out}
  {\bf{a}_{\rm out}}\left( {{\theta _2}} \right) = {{\bf{\Gamma }}_{{\theta _{1}},f}}{\bf{a}_{\rm in}}\left( {{\theta _1}} \right)
\end{equation}
where ${\bf{a}_{\rm out}}\left( {{\theta _2}} \right)$ can also be expressed as ${\left[
1, \cdots ,{{e^{-j\left( {N - 1} \right)\frac{{2\pi D}}{\lambda }\sin \left( {{\theta_2 }} \right)}}} \right]^T}$ which is the vector of the reflected EM wave towards $\theta_2$, ${{\bf{\Gamma }}_{{\theta _1},f}} = {\rm{diag}}\{ {e^{j{\Phi }\left( {{\theta _1},f,{C_1}} \right)}}, \cdots ,{e^{j{\Phi }\left( {{\theta _1},f,{C_N}} \right)}}\}  $ is the designed reflection coefficient matrix of RIS, where ${{\Phi }\left( {{\theta _1},f,{C_n}} \right)}$ is the reflection phase of the unit cells in the $n$-th column that tuned by varactor capacitance $C_n$. Without loss of generality, ${{\Phi }\left( {{\theta _1},f,{C_1}} \right)}=0$ can be assumed.\par

To realize the EM wave deflection, the designed phases should be arithmetic progression, i.e., ${{\Phi }\left( {{\theta _1},f,{C_n}} \right)}=\left(n-1\right){\Delta \Phi_{\theta_1,f}}$. When EM waves are reflected to $\theta_2$, as expressed in \eqref{eq:in_out}, the phase difference of adjacent unit cells ${\Delta \Phi_{\theta_1,f}}$ should satisfy
\begin{equation}\label{eq:recip_basic}
   - \frac{{2\pi D}}{\lambda }\sin \left( {{\theta _2}} \right) = \Delta {\Phi _{{\theta _1},f}} + \frac{{2\pi D}}{\lambda }\sin \left( {{\theta _1}} \right)+2k_{\theta_1}\pi,
\end{equation}
where {$k_{\theta_1}\in {\mathbb{Z}}$}.
Keeping the bias voltage network constant, another EM wave reversely impinges the RIS from $\theta_2$ and reflected to $\theta_3$ as shown in Fig. \ref{Fig.RIS} ($a$). Similarly, we have ${\bf{a}_{\rm out}}\left( {{\theta _3}} \right) = {{\bf{\Gamma }}_{{\theta _{2}},f}}{\bf{a}_{\rm in}}\left( {{\theta _2}} \right)$. In this reverse incident case, the phase difference of adjacent unit cells ${\Delta \Phi_{\theta_2,f}}$ should satisfy
\begin{equation}\label{eq:recip_basic_2}
   - \frac{{2\pi D}}{\lambda }\sin \left( {{\theta _3}} \right) = \Delta {\Phi _{{\theta _2},f}} + \frac{{2\pi D}}{\lambda }\sin \left( {{\theta _2}} \right)+2k_{\theta_2}\pi,
\end{equation}
where {$k_{\theta_2}\in{\mathbb{Z}}$}.
By substituting \eqref{eq:recip_basic} into \eqref{eq:recip_basic_2} and omitting $k_{\theta_1}$ and $k_{\theta_2}$ since the absolute difference between $\Delta {\Phi _{{\theta _1},f}} $ and $ \Delta {\Phi _{{\theta _2},f}}$ will not exceed $2\pi$, \eqref{eq:theta_3_1} can be proved.


\begin{small}

\end{small}

\end{document}